\documentclass[12pt]{elsarticle}

\usepackage{graphicx}
\usepackage{epstopdf}
\usepackage{lineno}
\usepackage{dcolumn}
\usepackage{bm}
\usepackage{amsmath}
\usepackage{amssymb}
\usepackage[latin1]{inputenc}
\usepackage[english]{babel}
\usepackage{subfigure}
\usepackage{float}

\makeatletter
\def\ps@pprintTitle{%
 \let\@oddhead\@empty
 \let\@evenhead\@empty
 \def\@oddfoot{}%
 \let\@evenfoot\@oddfoot}
 
 \newcommand\blfootnote[1]{%
  \begingroup
  \renewcommand\thefootnote{}\footnote{#1}%
  \addtocounter{footnote}{-1}%
  \endgroup
}

\begin{document}

\begin{frontmatter}

\title{Propagation of a Thermo-mechanical Perturbation on a Lipid Membrane}

\author{M. I. P\'erez-Camacho and J. C. Ruiz-Su\'arez$^{\ast}$}

\address{CINVESTAV-Monterrey, Autopista Nueva al Aeropuerto Km. 9.5, Apodaca, Nuevo Le\'on 66600, M\'exico}

\begin{abstract}
The propagation of sound waves on lipid monolayers supported on water has been studied during the melting transition. Since changes in volume, area, and compressibility in lipid membranes have biological relevance, the observed sound propagation is of paramount importance. However, it is unknown what would occur on a lipid bilayer, which is a more approximate model of a cell membrane. With the aim to answer this relevant question, we built an experimental setup to assemble long artificial lipid membranes. We found that if these membranes are heated in order to force local melting, a thermo-mechanical perturbation propagates a long distance. Our findings may support the existence of solitary waves, postulated to explain the propagation of isentropic signals together with the action potential in neurons.  
\end{abstract}

\end{frontmatter}

\blfootnote{\textit{$^{\ast}$E-mail: jcrs.mty@gmail.com}}

\section{Introduction}

Cells are enclosed by lipid membranes, complex structures responsible for giving integrity and controlling the flux of substances through them. Additionally, membranes are also the vehicle on which information propagates, for examples in specialized cells called neurons. Many reports have proposed different models of cell membranes to explain how this information, materialized in a signal called action potential, occurs and the function that its components have in the process. This is essential in the Hodgkin-Huxley model (H-H) \cite{hodkin}, which considers the action potential  as an electrical phenomenon induced by the flow of ions through protein channels, where the lipid membrane acts as a capacitor. Although the H-H model is the benchmark in neuroscience, there are some non-electrical events that cannot be explained in terms of ions flows. In nerves, the action potential is accompanied by mechanical displacements \cite{tasaki0, tasaki1, mamma} as well as adiabatic changes of temperature \cite{mamma, tasaki2, tasaki3, abbott, howart}. This, of course, is at odds with the fact the action potential is a dissipative process produced by the diffusion of ions through protein channels.

Moreover, it is well known that electrical, mechanical and thermodynamic properties are inherent of lipid membranes \cite{kauf1, kauf2} and a wide variety of biological systems exhibit spontaneous mechanical perturbations associated to electrical events \cite{mamma} and melting transitions below body temperature \cite{surfactante, soliton, soliton2}. The volume and area changes in the membranes have a close relationship with the pressure and temperature during the phase transition \cite{kay, liu, evans, evans2}. On fact, it has been reported that at the melting transition the average volume of a lipid membrane changes: in 1,2-dipalmitoyl-sn-glycerol-3-phosphocholine (DPPC) the volume change is near 4\%, while the lipid area varies 25\% \cite{nagle}. It is also known that the adiabatic compressibility, $k_s$, the heat capacity, $c_P$, and the permeability of the membrane display maxima in the melting regime \cite{soliton3, andrea}. Besides, under physiological conditions the membrane becomes softer when it is compressed and $k_s$ is frequency dependent \cite{mitaku}. For compressible fluids the speed of sound is given by $c=1/\sqrt{k_{s}\rho^A}$. Both, the dependence of $c$ on frequency and the increase of compressibility in the lipid melting transition are necessary conditions for the propagation of solitary pulses. Then, the presence of mechanical and reversible thermal changes during nerve transmission has led to an interesting suggestion: the action potential is an electromechanical pulse (or soliton) that travels along the lipid membrane and closely linked to the chain melting transition \cite{soliton}. Related to this idea, Heimburg and Jackson, based on previous works, proposed a model where any agent able to move the lipids through a phase transition could start a pulse. Indeed, phase transitions driven by a local change of temperature \cite{Spyropoulos}, a sudden change in pH \cite{trauble}, a local increase in calcium concentration \cite{calcio} or an electrostatic potential \cite{soliton2, kobatake}, could trigger this pulse.

The first study carried out to show the possibility of this phenomenon was done on a lipid monolayer \cite{2dpulse}, where the thermodynamic state is easily modified by a lateral pressure on a Langmuir balance \cite{poly}. It has been shown that in this system, acoustic pulses propagate triggered by small amounts of solvents released onto the monolayer surface \cite{2dpulse}. Also, the first evidence of solitary pulses on a lipid Langmuir monolayer and their dependence with the thermodynamic parameters was recently shown near the phase transition \cite{solitary, solitary2}. These works demonstrate the relationship between such pulses and the state of the lipid interface, agreeing with the soliton model originally proposed by Heimburg and Jackson. 

Nevertheless, from a biological perspective, a system that better resembles a real membrane is a lipid bilayer supported by a hydrophobic surface under water. The technique used to assemble such bilayers is called Black Lipid Membrane (BLM). In BLM, the diameter of the orifice where the bilayer is normally supported does not surpass 250 $\mu$m \cite{muller1}. Although large enough to study many electrical and elastic properties, such a short bilayer is inadequate to test the possible existence of a solitary wave. To circumvent this restriction, we designed an experimental set-up to assemble lipid membranes comparable in length to a real neuron (0.1 m).  Then, we carry out experiments where the membranes are locally heated within a tiny region. Far away from such region, we optically measured the arriving of a mechanical perturbation, presumably triggered by the thermal input. This observation gives further support to the idea that solitons do exist in real neurons. \\

\section{Materials and Methods}

N-decane, chloroform, methanol, n-hexadecane and n-pentane were obtained from Sigma Aldrich. 1,2-dipalmitoyl-sn-glycerol-3-phosphocholine \linebreak (DPPC) and 1,2-dimyristoyl-sn-glycero-3-phosphocholine (DMPC) were purchased from Avanti Polar Lipids (Birmingham, AL) and used without further purification. Lipid solutions (20mg/ml) in n-decane/chloroform/ methanol 7:2:1 were prepared and agitated for 20 minutes in order to achieve a homogeneous mixture. For all experiments purified water was used.

\begin{figure}[ht!]
\begin{center} 
\includegraphics[width=0.9\textwidth]{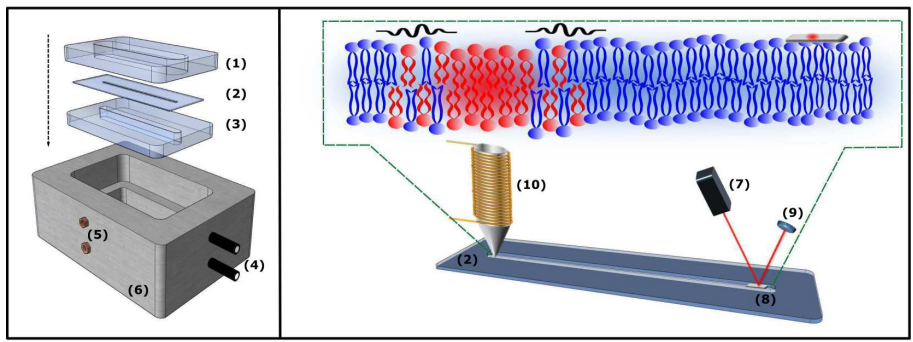}
\caption{\footnotesize (Color online)  Experimental set-up. (\textit{Left}) Schematic drawing of the chamber: (1) Acrylic top plate, (2) Acetate film with the aperture, (3) Acrylic bottom plate, (4) Water inlet valves, (5) Ag/AgCl electrodes connection, (6) Solid aluminium chamber. (\textit{Right}) Schematic of lipid membrane, heating and optical detector: (7) Light source, (8) Mylar target, (9) photodetector and (10) hot tip.}
\label{fig:camera} 
\end{center} 
\end{figure}


Figure \ref{fig:camera} depicts the lipid membrane system built to perform the experiments. It consists of an anodized aluminum block that is heated by circulating water (6). A system of parallel acrylic plates (1,3) is embedded in the block and it divides this block into two chambers. Each one has a water valve and an electrode connection (4,5). The linear lipid membrane is painted on a slit made on an acetate sheet (2). The slit  has 10.0 cm in length and 200 $\mu$m in width and is held in the middle by the acrylic plates. In each experiment the acetate was cleaned and then prepainted with 5\% n-hexadecane in n-pentane in order to reduce the mechanical tension at the edges of the aperture. The sandwich (acrylic plates and acetate) was carefully lowered inside the block. The chambers were filled with purified water. After the chosen temperature is reached and the system is in equilibrium, 20 mg/ml lipid solution in n-decane/chloroform/methanol 7:2:1 is deposited along the slit employing a micropipette. The Ag/AgCl electrodes were placed previously in each compartment and then connected to an electrometer (6517b, Keithley Instruments, Inc). The formation of membranes were controlled visually and  by electrical measurements with a triangular 100 mV voltage input pulse in order to verify the correct sealing of the membrane and the moment of its rupture.

Since the aim of this work was to propagate a mechanical perturbation along lipid membranes due to a phase transition, we first need to know their melting temperatures ($T_m$). DPPC and DMPC  have the following $T_m$'s: 41 and 24 $^\circ$C. However, the deposited mixtures contain solvents that do change these temperatures. In order to know their new values, we ought to perform a calorimetry analysis. Heat capacity profiles were recorded at a constant scan rate of 1$^\circ$ C/min and constant pressure of 3 atm. The lipid concentration was also 20 mg/ml dissolved in 7:2:1 n-decane/chloroform/methanol.  The calorimeter (Microcalorimeter, NanoDSC, TA Instrument) was interfaced to a PC, and data were analysed using the software provide with the instrument.  Then, the solution was dissolved in water (to induce the formation of vesicles). The suspension was stirred at 500 rpm for 30 min above the $T_m$. Finally, the sample was centrifuged and the supernatant was removed in order to eliminate the excess of solvent. For experiments without solvent, 3 mg/ml of lipid were hydrated with Milli-Q-water and stirred for 30 min at 50 $^\circ$C. In all samples, small unilammellar vesicles (SUVs) were subsequently obtained by extrusion with a polycarbonate filter pore size of 100 nm. Each experiment was equilibrated during 600 s at 2 $^\circ$C. A heating scan from 2 to 50 $^\circ$C was performed and it was followed by a cooling scan. This cycle was repeated 3 times for each sample.
 

Two main components conform the trigger/detection system separated from each other by a distance of 10 cm, see Figure \ref{fig:camera}: a heatable conical tip and an optical mirror where a laser beam is deflected. A hollow cone with a very sharp tip (diameter 100 microns) was machined at the end of an aluminium cylinder, which was coiled with a thin copper wire (10). The wire ends were connected to a voltage source. When the current flows, the wire heats and the tip of the cone is warmed up to a constant temperature measured by a thermocouple. The cylindrical cone is connected to a high precision micrometer in order for the tip to eventually touch the membrane with accuracy. In each experiment, the conical tip makes contact with the suspended membrane at one end of it, see right panel of Figure \ref{fig:camera}. Next, in order to measure a possible change of the lipid bilayer thickness at the inspection point, far away from the hot tip, an optical detector was used. After the bilayer is formed on the slit, a very tiny target of polyethylene terephthalate (reflective Mylar), weighing approximately 5 $\mu$g, was carefully placed onto the membrane (8). A laser beam (wavelength 632.8 nm) (7) is reflected on such mirror. A two segment photodiode (9) sensed the reflected beam, and this light signal produces a voltage, which is then amplified with a lock-in amplifier (HF2LI Lock-in Amplifier, Zurich Instruments AG) to improve the signal-to-noise ratio. Any perturbation of the membrane causes a change in the target position because the mirror can move freely.  If the mechanical perturbation is large enough, it throws the Mylar from its position, producing a signal voltage that cannot be recovered.


\section{Results}


\begin{figure}[ht!]
\begin{center} 
\includegraphics[width=0.7\textwidth]{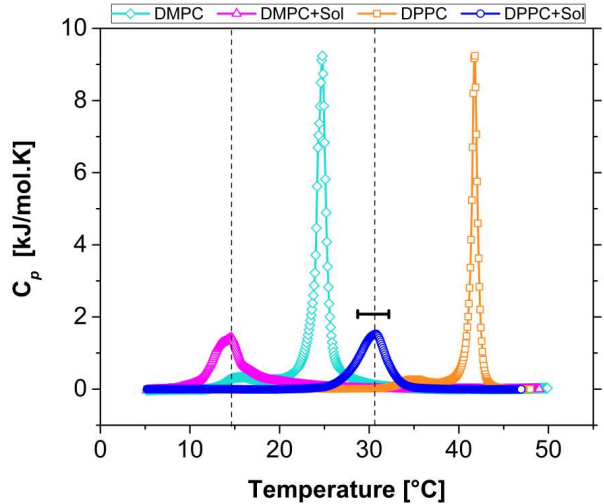}
\caption{\footnotesize (Color online) Calorimetric study of DPPC and DMPC liposomes and the effect of solvents in the calorimetric profiles. Yellow squares, control liposomes of DPPC at 3 mg/ml. Blue circles correspond to n-decane/chloroform/methanol in liposomes of DPPC. Cyan diamonds, control liposomes of DMPC at 3 mg/ml. Magenta triangles up show the effect of solvents in the temperature transition of liposomes of DMPC. Dashed lines mark the temperature transition shifts, which are approximately $|\Delta Tm|$=10 $^\circ$C.}
\label{fig:calorimetro} 
\end{center} 
\end{figure}



Figure \ref{fig:calorimetro} displays how the presence of solvents (in our case n-decane, chloroform, and methanol, necessary for providing stability in such large bilayers) induce shifts in the melting transition of membranes. Indeed, compared to the $T_m$'s of SUVs of pure lipids, the shifts are very large. For DPPC, the free solvent SUVs displays a principal peak at 41 $^\circ$C. The addition of the organic solvents during the process of preparation results in a rather small peak at a much lower temperature (30.7 $^\circ$C). Analogously, for solvent free DMPC vesicles $T_m$ is 24 $^\circ$C and once the solvents are added the peak attenuates and displaces to the left (14.6 $^\circ$C). Our results are in agreement with reports stating that the presence of short-chain alkanes in DPPC membranes decreases the melting temperature and broaden the transition \cite{mac, pope, decano, cloro1, cloro2}. In Figure \ref{fig:calorimetro}  the dashed lines mark the transition temperatures of the membranes with solvents.


\begin{figure}[ht!]
\begin{center} 
\includegraphics[width=0.7\textwidth]{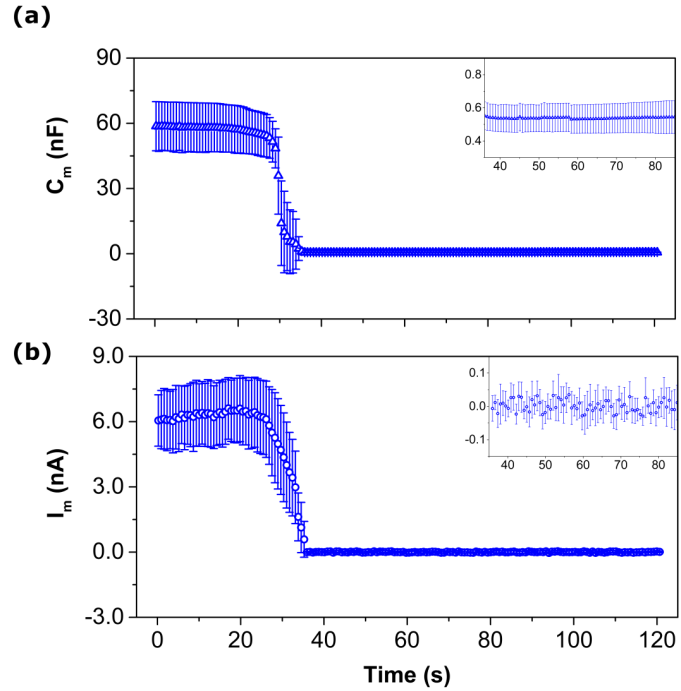}
\caption{\footnotesize (Color online) Electrical response of the system during lipid membrane painting at $T=25^\circ$C. Blue triangles correspond to capacitance measurement in a DPPC lipid bilayer. Blue circles show the effect of the DPPC lipid membrane in the current registered by the electrodes in the chamber. The insets show the capacitance and current values reached when the membrane is formed. Vertical lines mark the standard deviation for $n=5$.}
\label{fig:electrical} 
\end{center} 
\end{figure}



For reasons that we are going to clarify later, the temperature of the chamber was set at 25 $^\circ$C (for DPPC) and 35 $^\circ$C (for DMPC). Once the membranes are painted under water, the electrical capacitance and current display an important decrease. Figure \ref{fig:electrical} shows the electrical measurements for 5 different bilayers during their formation process. The amount of solvent during the painted produce differences in the width of bilayer displayed as a standard deviation. The formation of lipid bilayers is a spontaneous and self-assembly process.  It is very well known that the membrane formation is self-sealing process because a hole in a bilayer is energetically unfavorable \cite{berg}.

\begin{figure}[hb!]
\centering
 	\includegraphics[width=0.7\textwidth]{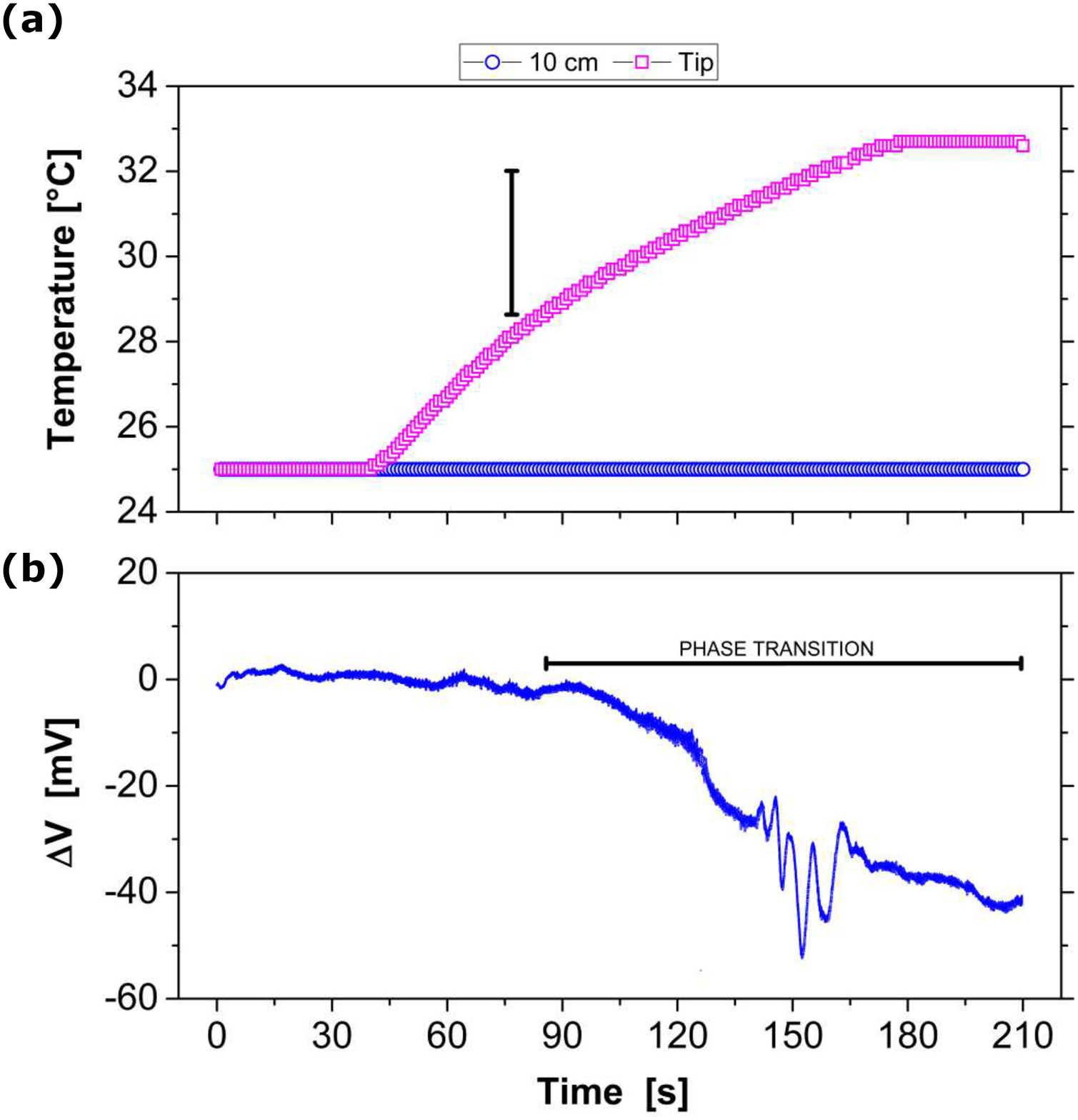}
 	\caption{\footnotesize (Color online)  Temperature and voltage measurements during the experiment. The black lines correspond to the range where the phase transition of the vesicles in the calorimeter occurs (see Figure \ref{fig:calorimetro}). (a) The temperature on the tip (submerged in water), as a function of time, is given by the magenta curve (squares). We note that it takes a bit more than two minutes for the coil to rise the temperature from the bath temperature to 33 $^\circ$C. The blue line (circles) corresponds to the temperature where the Mylar target  is positioned. (b) Representative voltage trace produced by the photodiode as a  function of time for a DPPC bilayer.  $\Delta V$ is the difference between the voltage output and its initial value $V_{0}$. The change in the voltage signal is produced during the phase transition.}
\label{fig:temp}  
\end{figure}

The stability of the lipid bilayers was then analysed, finding that they can last without rupture for 40 minutes, long enough to prepare and carry out the experiments. The tiny piece of mylar is gently deposited onto the membrane as observed in Figure \ref{fig:camera}. Thereafter the laser and detector are aligned, as assessed by the maximum voltage output $V_0$ given by the amplifier. Once the alignment is performed, the conical tip is lowered at the other end of the membrane until it makes contact with it. Fortunately for us the bilayer withstand and no rupture, as measured by the electrical current, is observed. The heater is turned on and the temperature of the metallic cone increases steadily, heating in turn the membrane just below the tip (as previously measured by a thermocouple), see magenta line (squares) in Figure \ref{fig:temp}(a). It is important to remark that the temperature where the piece of mylar was deposited (10 cm away from the hot tip), never senses an augment of temperature, see blue line (circles) in Figure \ref{fig:temp}(a).

\begin{figure}[ht!]
\centering
    \includegraphics[width=0.7\textwidth]{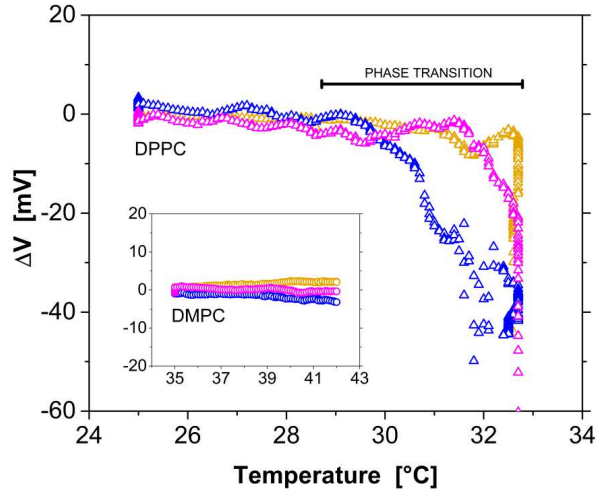}
    \caption{\footnotesize (Color online) Combining figures a and b in Figure \ref{fig:temp}  we obtain the voltage as a function of temperature on the tip.  Data for three experiments for DPPC are depicted. The black line corresponds to the phase transition observed in Figure \ref{fig:calorimetro}. Data for three experiments for DMPC are shown in the inset. See explanation in the main text.}
\label{fig:data2}
\end{figure}


As mentioned before, the temperature of the whole system (water and membrane) is 25 $^\circ$C (for DPPC). When the local temperature of the membrane (below the tip) reaches 28.7 $^\circ$C, two degrees below the transition temperature of DPPC/solvents depicted in Figure \ref{fig:calorimetro} (30.7 $^\circ$C), $\Delta V$ substantially descends as shown in Figure \ref{fig:temp}(b) and Figure \ref{fig:data2}. This means that the laser spot looses its alignment with the photodiode. Clearly, such misalignment must be caused by a change of position of the supported mylar. The fact this response is associated to the phase transition (marked by black lines in Figs. \ref{fig:calorimetro}-\ref{fig:data2}) hints towards the following episode: the membrane below the tip melts and this transition propagates along the membrane, reaching the position where the mylar is located. Indeed, lipids below and around the tip are forced to pass through a chain melting transition and both area per lipid and thickness of the bilayer change. For DPPC, the changes in area, volume and thickness at the transition are 24\%, 4.7\%, and -16\%, respectively \cite{soliton, soliton3}.

We now see the case of DMPC.  According to Figure \ref{fig:calorimetro}, its gel-fluid transition is at 14.6  $^\circ$C. Such low temperature allows us to use it as a control experiment. To stay well above the transition, this time we rise the temperature of the bath up to 35 $^\circ$C. Next, the membrane is locally heated with the conical tip from 35 to 42 $^\circ$C. In the inset of Figure \ref{fig:data2} we show the results for three experiments. Note that $\Delta V$ is constant, meaning that the Mylar is never perturbed. In other words, since the membrane remained at the same fluid phase, no expansion is produced. From this control experiment we also deduce that there are not a significant change in the volume of water or metal expansion of the tip during the heating that could affect the measurement.


\section{Discussion}

As mentioned in the introduction, evidence of mechanical changes in the membrane linked to the action potential was shown by Iwasa and Tasaki \cite{tasaki0, tasaki1}. The authors also discuss the relationship between the propagation of a nerve impulse with changes in the membrane temperature. The H-H model fails to explain these thermodynamic phenomena and therefore new points of view have emerged to explain the changes in the membrane during the action potential \cite{comparacion1, comparacion2}.

In artificial and biological membranes, the heat capacity and the compressibility are linked to the volume and area changes \cite{soliton3, surfactante}. The importance of changes in thermodynamic variables of the lipid bilayers during the phase transition has been explored in different works. A soliton model proposes the propagation of an isentropic pulse and its relation with mechanical dislocations, forces, voltage, and heat release in the membrane \cite{soliton}. In lipid membranes the transition at $T_m$ increases the compressibility, which also causes changes in the membrane thickness and lateral pressure. These are easily explored in lipid monolayers at the air-water interface and the existence of adiabatic sound waves has been studied  by Griesbauer et al. \cite{2dpulse, Griesbauer2009}. They report a sound velocity of surface waves derived from $k_{s}$ and $\rho^{A}$ as well as the dependence of propagation of pulses with the state of the lipid monolayer. This was also studied by Shrivastava et al \cite{dye}: the fluorescence intensity of dye molecules embedded in a lipid interface is sensitive to phase transitions. The same group has recently shown two-dimensional solitary elastic pulses in an air-water interface \cite{solitary}. In their experiments they show pulses propagating in the transition regime where the compressibility reaches a maximum. They varied the monolayer state by changing the surface density of lipid molecules, but it also can be modified by changing other physical features of the system. 

Despite the facility of working with lipid monolayers, they do not represent a real biological system. A bilayer system is a more approximate model. Moreover, in lipid bilayers the surface density and lateral pressure are intrinsic to the final state. We found in this work that such state can be locally modified by heat applied in a reduced zone of the bilayer. We suggest that as the lipids in this zone change from gel to fluid phase, a deformation is produced. This propagates along the membrane and eventually moves the laser target. How fast this perturbation moves is the next question we would like to answer in future work that is under progress, to hopefully confirm it behaves as a soliton. By the time being, we have learned here that a phase transition produced within a tiny region of a lipid membrane under water travels far away, suggesting that solitary waves in nerves may come to real terms.

This work has been supported by CONACYT, Mexico, under grants CB-220962 and FC-1132. 
M.I.P.C. acknowledges a scholarship from CONACYT. Fruitful discussions with Francisco Sierra and Victor Romero are
acknowledged.

\section*{References}

\bibliographystyle{unsrt}
{\scriptsize							
\bibliography{b1}}

\end{document}